\documentclass[12pt]{iopart}
\usepackage{iopams}  

\begin{document}

\eqnobysec
\newtheorem{proposition}{Proposition}[section]
\newtheorem{theorem}{Theorem}


\renewcommand\vec[1]{\boldsymbol{#1}}
\newcommand{\myset}[1]{\mathrm{#1}}
\newcommand{\mysetsize}[1]{|\myset{#1}|}
\newcommand\abel[1]{\mathfrak{a}\!\left(#1\right) }
\newcommand\myshiftedtheta[1]{\theta(\vec{z} + \abel{#1}) }
\newcommand\mymat[1]{\mathsf{#1}}  
\newcommand\transposed{^{\scriptscriptstyle t}} 
\def\myPhi(#1,#2,#3,#4){\Phi_{\scriptscriptstyle#2,#4}(#1,#3)}  
\def\myphi(#1,#2,#3,#4){\varphi_{\scriptscriptstyle#2,#4}(#1,#3)}  
\newcommand\myShifted[3][0]{\ifcase #1
   \mathbb{T}_{#2}{#3} \or                     
   \left(\mathbb{T}_{#2}{#3}\right) \else      
   #3[#2] 
   \fi}


\title{\\[16mm]Combinatorics of multisecant Fay identities.}
\author{V.E. Vekslerchik}
\address{
  Usikov Institute for Radiophysics and Electronics \\
  12, Proskura st., Kharkov, 61085, Ukraine 
}
\ead{vekslerchik@yahoo.com}


\begin{abstract}
We derive a set of identities for the theta functions on compact Riemann 
surfaces which generalize the famous trisecant Fay identity. Using these 
identities we obtain quasiperiodic solutions for a multidimensional 
generalization of the Hirota bilinear difference equation and for a 
multidimensional Toda-type system.
\end{abstract}


\section{Introduction.}

In the present work we derive some identities for the theta functions defined 
on the compact Riemann surfaces which generalize the famous Fay identity 
\cite{F73}. The classical trisecant Fay identity (TFI), 
see equation (45) from \cite{F73}, has been discovered as a result of the 
studies of the properties of the theta functions 
on abelian varieties and its proof is based on a rather complicated machinery 
of the algebraic geometry  \cite{F73,F85,G86,P92,M07}. 
The wide interest to the TFI stems, to a large extent, from the fact that it 
can be used to derive the quasiperiodic solutions for many integrable 
equations such as, for example, KdV, KP, sine-Gordon equations, Toda model etc. 
It has been shown that such solutions, previously obtained by the 
algebro-geometric approach (see, e.g., \cite{DMN76,K77a,K77b,K78}), 
naturally arise from this rather simple identity 
(see chapter IIIb of \cite{M07}). 

The aim of this work is to generalize the TFI bearing in mind its possible 
applications. 
We derive a set of identities that, as we hope, can be useful for obtaining 
solutions not only for the `classical' (1+1)-dimensional (like, for example 
the KdV, sine-Gordon or nonlinear Schr\"odinger equations) or (1+2)-dimensional 
models (like, for example, the KP, 2D Toda or Davey-Stewartson equations) 
but also for models in higher dimensions. 

In so doing, we do not rely on the algebraic geometry. 
After having formulated the TFI in section \ref{sec:tfi} we do not use any more 
properties of the Riemann surfaces, Abel mapping etc. 
As it turns out, even the explicit form of the coefficients that appear in the 
TFI is not important. What we need from the algebraic geometry are 1) 
existence of this identity and 2) its structure (see below). 
All calculations presented here are `elementary': we just `iterate' the TFI to 
obtain various consequences of this identity.

In section \ref{sec:phi} we introduce an auxiliary function, that takes into 
account the bilinearity of the TFI and `absorbs' most of the constants, and 
derive the main result of this paper (propositions \ref{prop:zax-th} and 
\ref{prop:zbx-th}) as well as a few more general bilinear identities. 
In section \ref{sec:multi} we go beyond the bilinear framework and obtain 
a set of multilinear identities. Since a vast area of practical use 
of the Fay-like identities is partial differential equations, we present 
in section \ref{sec:diff} some differential variants of the general identities. 
Finally, we give two examples of applications of the obtained results. 
In section \ref{sec:app} we derive quasiperiodic solutions for a 
multidimensional generalization of the famous Hirota bilinear difference 
equation and for a multidimensional Toda-type system.

\section{Trisecant Fay identity. \label{sec:tfi}} 

For a compact Riemann surface $\mathcal{X}$ of the genus $g$
one can introduce in a standard way 
a system of cuts $A_{i}$, $B_{i}$ ($i=1, ...,g$), 
a vector space of holomorphic 1-forms, 
its basis $\omega_{i}$, 
normalized by $\int_{A_{i}}\omega_{j} = \delta_{ij}$, 
the $g \times g$ complex matrix $\Omega$ 
with the elements $\int_{B_{i}}\omega_{j}$,
whose imaginary part is positive definite, 
the lattice 
$L_{\Omega} = \mathbb{Z}^{g} + \Omega \mathbb{Z}^{g}$ 
and the complex torus $\mathrm{Jac}(\mathcal{X}) = \mathbb{C}^{g}/L_{\Omega}$ 
(see, e.g., \cite{M07}). 
In this work, we intensively use the following fundamental constructions of the 
classical theory of compact Riemann surfaces: 
the theta function 
$\theta\left( \vec{z} \right) = \theta\left( \vec{z}, \Omega \right)$,
\begin{equation}
  \theta( \vec{z} ) = 
  \sum_{ \vec{n} \in \mathbb{Z}^{g} }
  \exp\left( 
    \pi i \, \vec{n}\transposed \Omega \vec{n}
    + 2 \pi i \, \vec{n}\transposed \vec{z}
  \right), 
\end{equation}
where $\vec{n}$ and $\vec{z}$ are $g$-column vectors, $\vec{n}\transposed$  
is a $g$-row vector (throughout this paper the symbol ${}\transposed$ 
stands for the transposition) 
and the Abel map $\mathcal{X} \to \mathrm{Jac}(\mathcal{X})$ is defined by 
\begin{equation}
\label{def:abel}
  \abel{x} = \int_{x_{0}}^{x} \vec{\omega} 
\end{equation}
where $\vec{\omega}$ is a column vector of the basis forms 
$\vec{\omega} = (\omega_{1}, ..., \omega_{g} )\transposed$ 
and $x_{0}$ is some fixed point of $\mathcal{X}$. The last definition can be 
extended to the definition of the Abel map from the space of divisors 
$\sum_{k} n_{k} x_{k}$ to $\mathrm{Jac}(\mathcal{X})$, 
\begin{equation}
  \abel{\sum_{k} n_{k} x_{k}} = \sum_{k} n_{k} \abel{x_{k}}, 
  \qquad
  n_{k} \in \mathbb{Z}, \;
  x_{k} \in \mathcal{X}.
\end{equation}

The aim of this work is to find generalizations of the TFI 
which we write as 
\begin{equation}
\label{id:fay}
\begin{array}{l}
    E(a,b) E(c,d) \; 
    \theta(\vec{z}) \, 
    \myshiftedtheta{a+b-c-d} 
\\ \qquad
  = 
    E(a,c) E(b,d) \; 
    \myshiftedtheta{a-d} \, 
    \myshiftedtheta{b-c}
\\ \qquad
  - E(a,d) E(b,c) \; 
    \myshiftedtheta{a-c} \, 
    \myshiftedtheta{b-d}. 
\end{array}
\end{equation}
where $a$, $b$, $c$ and $d$ are points of $\mathcal{X}$.  
The function $E(x,y)$ is a `scalar' version of the prime-form, 
\begin{equation}
\label{def:E}
	E(x,y) = 
	\theta\left[\!\! 
	  \begin{array}{c}
	  {\scriptscriptstyle\frac{1}{2}}\vec{m} \\ 
	  {\scriptscriptstyle\frac{1}{2}}\vec{n} 
	  \end{array}
	\!\!\right]
	(\abel{x-y})
  \qquad  
  \left( x,y \in \mathcal{X} \right) 
\end{equation}
where the theta function with characteristics is given by 
\begin{equation}
	\theta\Biggl[\!\! 
	  \begin{array}{c} \vec{a} \\ \vec{b} \end{array}
	\!\!\Biggr]
  ( \vec{z} ) 
  =
  \exp\left\{ \; 
      \pi i \, \vec{a}\transposed \Omega \vec{a} 
    + 2 \pi i \, \vec{a}\transposed \vec{b} 
    + 2 \pi i \, \vec{a}\transposed \vec{z} 
  \; \right\}
  \theta( \vec{z} + \Omega\vec{a} + \vec{b} ) 
\end{equation}
and $\vec{m}$, $\vec{n}$ are integer vectors,
$\vec{m},\vec{n} \in \mathbb{Z}^{g}$, related by 
\begin{equation}
  \vec{m}\transposed\vec{n} = \mbox{odd number}. 
\end{equation}  
Such choice of  $\vec{m}$ and $\vec{n}$ ensures the following properties of the 
function $E(x,y)$: 
\begin{equation}
  E(x,x) = 0, 
  \qquad
  E(x,y) = - E(y,x). 
\end{equation}

As has been mentioned in the Introduction, this paper is devoted to the 
`elementary' consequences of the Fay identity \eref{id:fay}, which means that 
starting from \eref{id:fay} we do not use the properties of the Riemann 
surfaces, Abel maps or other machinery of the algebraic geometry. 
For our purposes, even the form of $E(x,y)$ is not important 
(we present it just for the sake on completeness). 
What is important and what is repeatedly used in this work is that all 
coefficients in \eref{id:fay} are products of pairwise factors.

\section{$\Phi$-function and main identities. \label{sec:phi}}

Some part of the notation used in this paper deviates from the traditional 
algebro-geometrical one. So, for example, we almost do not use the notion of 
divisors (only as arguments of the Abel map $\abel{...}$). 
Instead, we prefer to formulate all results in terms of sets of the points of 
a Riemann surface (we often omit the words `of a Riemann surface').
Thus, instead of adding or subtracting divisors, we use the set operations with 
`+' and `$\backslash$' standing for the union and the difference of sets 
and $\mysetsize{...}$ for the number of elements of a set. 

It should be noted that throughout this paper we consider the 
`general position' case: 
there is no coinciding points in a set or, in other words, 
each point of a set appears there only once.

It turns out that calculations of this work turn out to be much more 
easy if performed not in terms of the theta functions, but in terms of the 
function $\Phi$ defined by 
\begin{equation}
  \Phi_{\myset{X},\myset{Y}}(\vec{z}, \myset{A},\myset{B}) 
  = 
  \myphi(\myset{A},\myset{X},\myset{B},\myset{Y}) 
  \frac{
    \myshiftedtheta{\myset{X} \backslash \myset{A}{+}\myset{B}} 
    \myshiftedtheta{\myset{A} {+} \myset{Y} \backslash \myset{B}} 
  }{
    \myshiftedtheta{\myset{X}} 
    \myshiftedtheta{\myset{Y}} 
  }
\end{equation}
with the constants (in the sense that they do not depend on $\vec{z}$) 
\begin{equation}
\label{def:phi}
  \myphi(\myset{A},\myset{X},\myset{B},\myset{Y}) 
  = 
  \frac{ E(\myset{B},\myset{X}\backslash\myset{A}) 
         E(\myset{A},\myset{Y}\backslash\myset{B}) }
       { E(\myset{A},\myset{X}\backslash\myset{A}) 
         E(\myset{B},\myset{Y}\backslash\myset{B}) }
\end{equation}
where 
\begin{equation}
  E(\myset{X},\myset{Y}) 
  = 
  \prod_{x \in \myset{X}} \prod_{y \in \myset{Y}} E(x,y). 
\end{equation}
In the following formulae we usually do not indicate the dependence 
on $\vec{z}$ explicitly: 
we consider $\vec{z}$ being fixed and write 
$\myPhi(\myset{A},\myset{X}, \myset{B},\myset{Y})$ 
instead of  
$\Phi_{\myset{X}\myset{Y}}(\vec{z}, \myset{A},\myset{B})$.

In terms of $\Phi$, one can rewrite the Fay identity \eref{id:fay} in 
different ways:
\begin{equation}
\label{id:zax31}
     \myPhi(a,\myset{X},\emptyset,\myset{Y}) 
  +  \myPhi(b,\myset{X},\emptyset,\myset{Y}) 
  +  \myPhi(c,\myset{X},\emptyset,\myset{Y}) 
  = 0,
  \quad
  \myset{X}= \{a,b,c\}, \; \myset{Y}= \{d\}, 
\end{equation}
or 
\begin{equation}
\label{id:zax22c}
  \myPhi(a,\myset{X},c,\myset{Y}) +  \myPhi(b,\myset{X},c,\myset{Y}) 
  = 1, 
  \qquad
  X= \{a,b\}, \; Y= \{c,d\}.
\end{equation}
Hereafter we do distinguish between 1-point sets and points of $\mathcal{X}$ and write 
$a$ instead of $\{a\}$ etc. 
These formulae not only look more simple than the original one, 
but also reveal some inner structures behind the TFI. 
And indeed, identities \eref{id:zax31} and \eref{id:zax22c} can be generalized 
to the case of arbitrary sets $\myset{X}$ and $\myset{Y}$ to become the 
multisecant Fay identities.

\begin{proposition} \label{prop:zax}

For arbitrary sets $\myset{X}$ and $\myset{Y}$ related by 
$ \mysetsize{X} = |\myset{Y}| + 2$ 
the function $\Phi$ satisfies 
\begin{equation}
\label{id:zax}
  \sum_{x \in \myset{X}} \myPhi(x,\myset{X},\emptyset,\myset{Y}) = 0.
\end{equation}

\end{proposition} 

\begin{proposition} \label{prop:zbx}

For arbitrary sets $\myset{X}$ and $\myset{Y}$ related by 
$\mysetsize{X} = \mysetsize{Y}$ 
the function $\Phi$ satisfies 
\begin{equation}
\label{id:zbx}
  \sum_{x \in \myset{X}} \myPhi(x,\myset{X},y,\myset{Y})  
  = 
  1,
  \qquad
  \forall y \in \myset{Y}, 
\end{equation}
\begin{equation}
\label{id:zby}
  \sum_{y \in \myset{Y}}  \myPhi(x,\myset{X},y,\myset{Y})  
  = 
  1,
  \qquad
  \forall x \in \myset{X}. 
\end{equation}

\end{proposition} 
%
We present proofs of these results in \ref{proof:zax} and \ref{proof:zbx}.

A simple consequence of, for example, \eref{id:zby} can be 
obtained by noting that it holds for any choice of $x$ among the points of the 
set $\myset{X}$. Thus, multiplying \eref{id:zby} by arbitrary 
constant $\Gamma_{x}$ and summarizing over $\myset{X}$ leads to 
\begin{equation}
  \sum_{x \in \myset{X}}  
  \sum_{y \in \myset{Y}}  
  \Gamma_{x} \myPhi(x,\myset{X},y,\myset{Y})  
  = 
  \sum_{x \in \myset{X}}  \Gamma_{x},
  \qquad 
  (\mysetsize{X} = \mysetsize{Y}). 
\end{equation}
Thus one can convert the inhomogeneous identities \eref{id:zbx} and 
\eref{id:zby} into homogeneous ones by imposing the restriction 
$\sum_{x \in \myset{X}}  \Gamma_{x} = 0$.

Before proceed further, we would like to rewrite the obtained 
identities in the `original' theta-form.

\begin{proposition} \label{prop:zax-th} 

For all $\vec{z}$ and sets $\myset{X}$ and $\myset{Y}$ related by 
$ \mysetsize{X} = \mysetsize{Y} + 2 $ the theta function satisfies 
\begin{equation}
\label{id:zax-th}
  \sum_{x \in \myset{X}}   
  \varphi_{\scriptscriptstyle \myset{X},\myset{Y}}(x) \; 
  \myshiftedtheta{\myset{X}\backslash x} \, 
  \myshiftedtheta{\myset{Y}{+}x} 
  = 0
\end{equation}
with 
\begin{equation}
  \varphi_{\scriptscriptstyle \myset{X},\myset{Y}}(x) =    
  \frac{ E(x,\myset{Y}) }{  E(x,\myset{X}\backslash x) }. 
\end{equation}

\end{proposition} 

\begin{proposition} \label{prop:zbx-th} 

For all $\vec{z}$ and sets $\myset{X}$ and $\myset{Y}$ related by 
$\mysetsize{X} = \mysetsize{Y}$ the theta functions satisfies 
\begin{equation}
\label{id:zbx-th}
\fl
  \myshiftedtheta{\myset{X}} \, 
  \myshiftedtheta{\myset{Y}} 
  = 
  \sum_{x \in \myset{X}} 
    \myphi(x,\myset{X},y,\myset{Y}) \, 
    \myshiftedtheta{\myset{X} \backslash x{+}y} \, 
    \myshiftedtheta{\myset{Y} \backslash y{+}x} 
\end{equation}
for any $y \in \myset{Y}$ with 
\begin{equation}
  \myphi(x,\myset{X},y,\myset{Y}) 
  = 
  \frac{ E(x,\myset{Y}\backslash y) E(y,\myset{X}\backslash x) }
       { E(x,\myset{X}\backslash x) E(y,\myset{Y}\backslash y) }.
\end{equation}

\end{proposition} 

\bigskip

The next step in generalizing the Fay identities can be done by switching 
from summation over the points of a set to summation over subsets of a given 
set of points. To make the following formulae more legible we will use the 
subscript to indicate the size of a set:
\begin{equation}
  \myset{X}_{n} = \left\{ x_{1}, \, ... \, , x_{n} \right\}. 
\end{equation}
With this change of the notation, we can formulate the following results.

\begin{proposition} \label{prop:ZAX}
For arbitrary sets $\myset{X}_{n+2}$ and $\myset{Y}_{n}$ and any 
$m \in [0,n]$
\begin{equation}
\label{id:ZAX} 
  \sum_{ \myset{A}_{m+1} \subset \myset{X}_{n+2}} 
  \myPhi(\myset{A}_{m+1},\myset{X}_{n+2},\myset{B}_{m},\myset{Y}_{n})  
  = 
  0, 
  \qquad
  \myset{B}_{m} \subset \myset{Y}_{n} 
\end{equation}

\end{proposition} 

\begin{proposition} \label{prop:ZBX}
For arbitrary sets $\myset{X}_{n}$ and $\myset{Y}_{n}$ and any 
$m \in [1,n]$
\begin{equation}
\label{id:ZBX} 
  \sum_{\myset{A}_{m} \subset \myset{X}_{n}} 
  \myPhi(\myset{A}_{m},\myset{X}_{n},\myset{B}_{m},\myset{Y}_{n})  
  = 
  1,
  \qquad 
  \myset{B}_{m} \subset \myset{Y}_{n} 
\end{equation}

\end{proposition} 
%
We present proofs of these results in \ref{proof:ZAX} and \ref{proof:ZBX}.

In terms of the theta functions, identities \eref{id:ZAX} and  \eref{id:ZBX} 
read
\begin{eqnarray}
&&
\fl
  \sum_{\myset{A}_{m+1} \subset \myset{X}_{n+2}} 
  \myphi(\myset{A}_{m+1},\myset{X}_{n+2},\myset{B}_{m},\myset{Y}_{n}) \, 
  \myshiftedtheta{ \myset{X}_{n+2}\backslash\myset{A}_{m+1}{+}\myset{B}_{m} } \, 
  \myshiftedtheta{ \myset{A}_{m+1}{+}\myset{Y}_{n}\backslash\myset{B}_{m} }
\nonumber\\&&
  = 0
\\[2mm]&&
\fl
  \sum_{\myset{A}_{m} \subset \myset{X}_{n}} 
  \myphi(\myset{A}_{m},\myset{X_{n}},\myset{B}_{m},\myset{Y}_{n}) \, 
  \myshiftedtheta{\myset{X}_{n}\backslash\myset{A}_{m}{+}\myset{B}_{m}} \, 
  \myshiftedtheta{\myset{A}_{m}{+}\myset{Y}_{n}\backslash\myset{B}_{m}} 
\nonumber\\&&
  = 
  \myshiftedtheta{\myset{X}} 
  \myshiftedtheta{\myset{Y}} 
\end{eqnarray}

\section{Multilinear Fay identities. \label{sec:multi}}

In this section we rewrite some of the obtained identities in the matrix form 
and derive, using this representation, various multilinear ones.

For given sets $\myset{X}$ and $\myset{Y}$ of equal size $n$, 
\begin{equation}
  \myset{X} = \{ x_{1}, \, ... \, , x_{n} \},
  \quad
  \myset{Y} = \{ y_{1}, \, ... \, , y_{n} \},
\end{equation}
consider the $(n \times n)$-matrix 
\begin{equation}
  \mymat{\Phi}_{\myset{X}\myset{Y}} 
  = 
  \biggl( \myPhi(x_{j},X,y_{k},Y) \biggr)_{j,k = 1, ... , n}
\end{equation}
In terms of $\mymat{\Phi}_{\myset{X}\myset{Y}}$, identities \eref{id:zbx} and 
\eref{id:zby} become 
\begin{equation}
\label{eq:Pu}
  \begin{array}{l}
	\vec{u}\transposed \mymat{\Phi}_{\myset{X}\myset{Y}} 
	= 
	\vec{u}\transposed, 
	\\[2mm]
	\mymat{\Phi}_{\myset{X}\myset{Y}} \vec{u} 
	= 
	\vec{u} 
	\end{array}
\end{equation}
where $\vec{u}$ is the $n$-column with all components equal to $1$, 
$ \vec{u} = (1, \, ... \, ,1 )\transposed $. 

One can easily `iterate' these formulae to obtain more complex ones. 
For example, multiplication (from the right-hand side) by 
$\mymat{\Phi}_{\myset{Y}\myset{Z}}$, where $\mysetsize{Z}=n$, leads to 
\begin{equation}
\label{eq:uPP}
	\vec{u}\transposed 
	\mymat{\Phi}_{\myset{X}\myset{Y}} 
	\mymat{\Phi}_{\myset{Y}\myset{Z}}
	= 
	\vec{u}\transposed 
	\mymat{\Phi}_{\myset{Y}\myset{Z}}
	= 
	\vec{u}\transposed. 
\end{equation}
In as similar way one can obtain
\begin{equation}
  \begin{array}{l}
	\vec{u}\transposed 
	\mymat{\Phi}_{\myset{X}\myset{U}_1} 
  \mymat{\Phi}_{\myset{U}_{1}\myset{U}_{2}} 
	...
	\mymat{\Phi}_{\myset{U}_{l-1}\myset{U}_{l}} 
	\mymat{\Phi}_{\myset{U}_{l}\myset{Y}} 
	= 
	\vec{u}\transposed, 
	\\[2mm]
	\mymat{\Phi}_{\myset{X}\myset{U}_{1}} 
  \mymat{\Phi}_{\myset{U}_{1}\myset{U}_{2}} 
	...
  \mymat{\Phi}_{\myset{U}_{l-1}\myset{U}_{l}} 
	\mymat{\Phi}_{\myset{U}_{l}\myset{Y}} 
	\vec{u}
	= 
	\vec{u} 
	\end{array}
\label{eq:mlm}
\end{equation}
($\mysetsize{U_{1}} = ... = \mysetsize{U_{l}} = n$).

To return to the standard, `scalar', identities one can use an arbitrary 
vector $\vec{v} \in \mathbb{C}^{n}$ which yields
\begin{equation}
	\vec{u}\transposed 
	\mymat{\Phi}_{\myset{X}\myset{U}_1} 
  \mymat{\Phi}_{\myset{U}_{1}\myset{U}_{2}} 
	...
	\mymat{\Phi}_{\myset{U}_{l-1}\myset{U}_{l}} 
	\mymat{\Phi}_{\myset{U}_{l}\myset{Y}} 
	\vec{v} 
	= 
	\vec{u}\transposed 
	\vec{v} 
\label{eq:mls}
\end{equation}
Depending on the choice of $\vec{v}$ one can arrive at 
the homogeneous identities 
(if $\vec{u}\transposed \vec{v} = 0$) 
or at the inhomogeneous ones
(if $\vec{u}\transposed \vec{v} \ne 0$). 

The key moment is that all identities discussed in the previous section 
were bilinear in $\theta$, like the original Fay identity \eref{id:fay}. 
At the same time, in \eref{eq:mlm} or \eref{eq:mls} we have products 
of $l+1$ bilinear in $\theta$ matrices. 
This means that we have derived, by elementary calculations, a large set of 
multilinear Fay identities.

It is easy to see that all above calculations can be repeated in the case of 
rectangular matrices $\mymat{\Phi}_{\myset{X}\myset{Y}}$, i.e. one can lift the 
condition $\mysetsize{X}=\mysetsize{Y}$. However, we restrict ourselves with 
the simplest case. 

Another way to obtain the multilinear Fay identities is to consider 
determinants that appear in the matrix identities presented above. 
For example, equation \eref{eq:Pu} states that $\vec{u}$ is the eigenvector of 
$\mymat{\Phi}_{\myset{X}\myset{Y}}$ corresponding to the unit eigenvalue, which 
leads to 
\begin{equation}
\label{eq:det-phi}
  \det\left| \mymat{\Phi}_{\myset{X}\myset{Y}} - \mymat{1} \right| 
  = 0.
\end{equation} 
In a similar way, equations \eref{eq:mlm} imply 
\begin{equation}
  \det\left| 
  \mymat{\Phi}_{\myset{X}\myset{U}_1} 
  \mymat{\Phi}_{\myset{U}_{1}\myset{U}_{2}} 
	...
	\mymat{\Phi}_{\myset{U}_{l-1}\myset{U}_{l}} 
	\mymat{\Phi}_{\myset{U}_{l}\myset{Y}} 
	- 
	\mymat{1} 
  \right| 
  = 0. 
\end{equation}
As another example, one can note that equation \eref{eq:uPP} implies that 
$
\mymat{\Phi}_{\myset{X}\myset{Y}}\mymat{\Phi}_{\myset{Y}\myset{Z}} 
- \mymat{\Phi}_{\myset{X}\myset{Z}}
$ 
is a degenerate matrix (it sends the row $\vec{u}\transposed$ to zero), 
which leads to 
\begin{equation}
  \det\left| 
  \mymat{\Phi}_{\myset{X}\myset{Y}}
  \mymat{\Phi}_{\myset{Y}\myset{Z}} 
  - 
  \mymat{\Phi}_{\myset{X}\myset{Z}}
  \right| 
  = 0 
\end{equation}
with obvious generalization to the different products of the matrices 
similar to ones that appear in \eref{eq:mlm}.

Note that \eref{eq:det-phi} and other determinant identities differ from 
the determinant identity derived by Fay (see equation (43) in \cite{F73}).

\section{Differential Fay identities. \label{sec:diff}} 

For two close points $p$ and $q$ of a Riemann surface $\mathcal{X}$, there 
naturally appear two `small' (i.e. vanishing when $p \to q$) quantities 
\begin{equation}
	\vec{\delta}_{pq} = \abel{p-q} 
	\in \mathrm{Jac}(\mathcal{X}) 
\end{equation}
and 
\begin{equation}
	\varepsilon_{pq} = E(p,q) 
	\in \mathbb{C}.
\end{equation}
After introducing the differential operator $\partial_{q}$ by 
\begin{equation}
	\partial_{q} \theta(\vec{z}) 
	= 
	\lim_{p \to q} 
	\frac{1}{ \varepsilon_{pq} }
	\left[ \theta(\vec{z} + \vec{\delta}_{pq}) - \theta(\vec{z}) \right]
\end{equation}
and defining the constant $\Lambda_{q,x,y}$ as 
\begin{equation}
	\Lambda_{q,x,y} = 
	\frac{ 1 }{ E(q,x)E(q,y) }
	\lim_{p \to q} 
	\frac{1}{ \varepsilon_{pq} }
	\left[ E(p,x)E(q,y) - E(q,x)E(q,y) \right]
\end{equation}
one can obtain from the Fay identity \eref{id:fay} 
\begin{equation}
\label{eq:dt}
\begin{array}{l} 
	\left[ D_{q} + \Lambda_{q,x,y} \right] 
	\theta(\vec{z} + \abel{x-y}) \cdot \theta(\vec{z}) 
\\ \qquad
	= 
	\frac{ \displaystyle E(x,y) }{ \displaystyle E(x,q) E(y,q) } \; 
	\myshiftedtheta{x-q} \myshiftedtheta{q-y} 
\end{array}
\end{equation}
where $D_{q}$ is the Hirota bilinear operator, 
\begin{equation}
	D_{q} \, u \cdot v 
	= 
	\left( \partial_{q} u \right) v 
	- 
	u \left( \partial_{q} v \right).
\end{equation}
Similar calculations, starting from \eref{id:zax-th} with $\myset{X}$ 
replaced with $\myset{X} + p + q$, lead to the following generalization of 
\eref{eq:dt}.

\begin{proposition} 

For all $\vec{z}$ and sets $\myset{X}$ and $\myset{Y}$ related by 
$ \mysetsize{X} = \mysetsize{Y}$ 
but arbitrary otherwise the theta function satisfies 
\begin{equation}
\begin{array}{l}
	\left[ D_{q} + \Lambda_{q,X,Y} \right] 
  \myshiftedtheta{\myset{X}} \cdot \myshiftedtheta{\myset{Y}} 
\\[2mm] \qquad
  = 
  \sum\limits_{x \in \myset{X}}  
  \psi_{q,\myset{X},\myset{Y}}(x) \, 
  \myshiftedtheta{\myset{X}\backslash x{+}q} 
  \myshiftedtheta{\myset{Y}{+}x{-}q} 
\\[2mm] \qquad
  = 
  -
  \sum\limits_{y \in \myset{X}}  
  \psi_{q,\myset{Y},\myset{X}}(y) \, 
  \myshiftedtheta{\myset{X}{+}y{-}q} 
  \myshiftedtheta{\myset{Y}\backslash y{+}q}. 
\end{array}
\end{equation}
where 
\begin{equation}
	\psi_{q,\myset{X},\myset{Y}}(x) = 
	\frac{ E(q,\myset{X}) E(x,\myset{Y}) } 
       { E(q,\myset{Y}) E^{2}(q,x) E(x,\myset{X}\backslash x) } 
\end{equation}
and 
\begin{equation}
	\Lambda_{q,\myset{X},\myset{Y}} = 
	\frac{ 1 }{ E(q,\myset{X})E(q,\myset{Y}) }
	\lim_{p \to q} 
	\frac{1}{ \varepsilon_{pq} }
	\left[ E(p,\myset{X})E(q,\myset{Y}) - E(q,\myset{X})E(q,\myset{Y}) \right]. 
\end{equation}

\end{proposition} 

\section{Applications. \label{sec:app}} 

In this section we would like to discuss the `practical' aspects of the 
obtained results. Our aim is to show how one can use the theta functions to 
derive solutions for multidimensional versions of the well-known integrable 
models.

\subsection{
The $n$-dimensional version of the Hirota bilinear discrete equation. 
\label{sec:hbde}} 

Let us return to the equation \eref{id:zax-th}, 
\begin{equation}
  \sum_{x \in \myset{X}}   
  \varphi_{\myset{X},\myset{Y}}(x) \; 
  \myshiftedtheta{\myset{X}\backslash x}
  \myshiftedtheta{\myset{Y}{+}x} 
  = 0
  \quad
  (\mysetsize{X} = \mysetsize{Y} + 2) 
\end{equation}
for
\begin{equation}
  \myset{X} = \{ x_{1}, \, ... \, , x_{n} \},
\end{equation}
which, after the shift 
$\vec{z} \to \vec{z} - \frac{1}{2} \abel{\myset{X}{+}\myset{Y}}$ 
can be rewritten as 
\begin{equation}
  \sum_{k=1}^{n}   
  \Gamma_{k} \; 
  \theta\left(\vec{z} + \vec{e}_{k} \right) 
  \theta\left(\vec{z} - \vec{e}_{k} \right) 
  = 0
\end{equation}
where  
\begin{equation}
\label{eq-hbde-G}
  \Gamma_{k} = \varphi_{\myset{X},\myset{Y}}(x_{k}), 
  \qquad 
  \vec{e}_{k} 
  = 
  \abel{x_{k}} 
  + {\scriptstyle \frac{1}{2}} \abel{\myset{Y}} 
  - {\scriptstyle \frac{1}{2}} \abel{\myset{X}}. 
\end{equation}
It is easy to see that this equation implies that the function 
\begin{equation}
  \Theta(m_{1}, \, ... \, , m_{n} )
  = 
  \theta\left( \vec{z} + \sum_{k=1}^{n} m_{k} \vec{e}_{k} \right) 
\end{equation}
satisfies the equation 
\begin{equation}
  \sum_{k=1}^{n} 
  \Gamma_{k} 
  \Theta( \, ... \, , m_{k}+1 , \, ... \, )
  \Theta( \, ... \, , m_{k}-1 , \, ... \, ) 
  = 0, 
\end{equation}
which is similar to the n-dimensional Hirota bilinear discrete equation,
\begin{equation}
  \sum_{k=1}^{n} 
  \tau( \, ... \, , m_{k}+1 , \, ... \, )
  \tau( \, ... \, , m_{k}-1 , \, ... \, ) 
  = 0,
\end{equation}
but with extra coefficients $\Gamma_{k}$. 
One can take into account this difference, by introducing the 
quadratic in $m_{k}$ function 
\begin{equation}
\label{eq-hbde-f}
  f( m_{1}, \, ... \, , m_{n} ) 
  =
  \frac{1}{2} 
  \sum_{k=1}^{n} m_{k}^{2} \ln\Gamma_{k}. 
\end{equation}
To summarize, we can state the following result.

\begin{proposition} \label{prop:hbde}

For arbitrary vector $\vec{z}$, $n$-set $\myset{X}$ and $(n-2)$-set $\myset{Y}$ 
equations \eref{eq-hbde-G} and \eref{eq-hbde-f} determine a solution 
\begin{equation}
  \tau( m_{1}, \, ... \, , m_{n} ) 
  =
  \exp\left[ f( m_{1}, \, ... \, , m_{n} ) \right] 
  \theta\left( \vec{z} + \sum_{k=1}^{n} m_{k} \vec{e}_{k} \right) 
\end{equation}
for the $n$-dimensional version of the Hirota bilinear discrete equation 
\begin{equation}
  \sum_{k=1}^{n} 
  \tau( \, ... \, , m_{k}+1 , \, ... \, )
  \tau( \, ... \, , m_{k}-1 , \, ... \, ) 
  = 0.
\end{equation}

\end{proposition} 

\subsection{The $n$-dimensional Toda-type lattice. 
} 

Consider the situation when $\mysetsize{X}=\mysetsize{Y}=n$ and 
\begin{equation}
\label{cond:toda}
  \abel{\myset{X}} = \abel{\myset{Y}}. 
\end{equation}
In this case equation \eref{id:zbx-th}, after the shift 
$\vec{z} \to \vec{z} - \abel{\myset{X}}$, can be written as  
\begin{equation}
  \theta^{2}(\vec{z}) 
  = 
  \sum_{k=1}^{n}   
  \Gamma_{k} \; 
  \theta(\vec{z} + \vec{e}_{k}) 
  \theta(\vec{z} - \vec{e}_{k}) 
\end{equation}
where  
\begin{equation}
\label{eq-toda-G}
  \Gamma_{k} = \myphi(x_{k},\myset{X},y,\myset{Y}), 
  \qquad 
  \vec{e}_{k} 
  = 
  \abel{x_{k}} - \abel{y}. 
\end{equation}
Eliminating the constants $\Gamma_{k}$ and making obvious definitions 
we can formulate the following result.

\begin{proposition} \label{prop:toda}

For arbitrary vector $\vec{z}$ and two $n$-sets $\myset{X}$ and $\myset{Y}$ 
related by $\abel{\myset{X}} = \abel{\myset{Y}}$ function  
\begin{equation}
  u( m_{1}, \, ... \, , m_{n} ) 
  =
  f( m_{1}, \, ... \, , m_{n} )  
  + 
  \ln\theta\left( \vec{z} + \sum_{k=1}^{n} m_{k} \vec{e}_{k} \right), 
\end{equation}
where 
\begin{equation}
  f( m_{1}, \, ... \, , m_{n} ) 
  = 
  \frac{1}{2} 
  \sum_{k=1}^{n} m_{k}^{2} \, \ln\Gamma_{k} 
\end{equation}
with $\vec{e}_{k}$ and $\Gamma_{k}$ defined in \eref{eq-toda-G}, 
satisfies the $n$-dimensional Toda-type equation
\begin{equation}
  \sum_{k=1}^{n} 
  \exp\left( \Delta_{k} u \right) 
  = 1
\end{equation}
where the second-order difference operators $\Delta_{k}$ are defined by 
\begin{equation}
\fl
  (\Delta_{k} F)(m_{1}, \, ... \, m_{n}) 
  = 
  F( \, ... \, , m_{k}+1 , \, ... \, )
  - 2 F( \, ... \, , m_{k}, \, ... \, )
  + F( \, ... \, , m_{k}-1 , \, ... \, ). 
\end{equation}

\end{proposition} 

\section{Discussion.} 

In this paper we have presented a number of identities for the theta functions 
defined on the compact Riemann surfaces which generalize the TFI. 
We would like to repeat that all these identities were obtained by iteration of 
the original TFI \eref{id:fay} without using any additional facts from the 
algebraic geometry. 

The main idea behind this work is to facilitate usage of the multidimensional 
theta functions in the applied problems like solving the differential or 
difference equations. We hope that the approach of this work and the obtained 
results give possibility to address these questions in a more easy way, 
without necessity to develop each time the algebro-geometric scheme, 
involving, for example, Baker-Akhiezer functions, Riemann-Roch theorem etc 
like, e.g., in \cite{DMN76,K77a,K77b,K78}.

We are aware of the fact that the presented identities should be discussed 
from the viewpoint of the algebraic geometry. For example, we have used the 
conditions like \eref{cond:toda} but did not mention their meaning 
from the angle of the existence of meromorphic functions with given principal 
divisors. As another example, we never tried to interpret functions like 
\eref{def:phi} as a generalized cross-ratios. 
Such questions are important not only from the viewpoint of unification of 
different approaches. For example, when dealing with large sets of points 
(in our terms), whose size is greater than the genus $g$ of the Riemann 
surface, one has to consider the possibility of trivialization of some of the 
identities. 
However, these questions are out of the scope of the present paper and may be 
addressed in separate studies.

\appendix

\section{Proof of proposition \ref{prop:zax}. \label{proof:zax}} 

Consider the expression that appears in the left-hand side of \eref{id:zax},
\begin{equation}
\label{def:zax}
  \mathfrak{o}_{\myset{X},\myset{Y}} 
  = 
  \sum_{x \in \myset{X}} \myPhi(x,\myset{X},\emptyset,\myset{Y}).  
\end{equation}
Using equation \eref{id:fay} with $a=x_1$, $b=x_2$, $c=x$ and $d=y_1$ 
shifted by $\myset{X}+y_1$
one can present the summand in the last equation as 
\begin{equation}
\begin{array}{ll}
  \myPhi(x,\myset{X}+x_{1}+x_{2},\emptyset,\myset{Y}+y_{1}+y_{2}) 
  &
  = 
  \myShifted[1]{\myset{X}}{ \myPhi(x_{1},x_{1}+x_{2},\emptyset,y_{1})} 
  \myShifted[1]{y_{1}}{ \myPhi(x,\myset{X}+x_{1},\emptyset,\myset{Y}+y_{2})} 
  \\&
  + 
  \myShifted[1]{\myset{X}}{ \myPhi(x_{2},x_{1}+x_{2},\emptyset,y_{1})} 
  \myShifted[1]{y_{1}}{ \myPhi(x,\myset{X}+x_{2},\emptyset,\myset{Y}+y_{2})} \\[1mm]
\end{array}
\end{equation}
which, together with 
\begin{equation}
  \myShifted[1]{\myset{X}}{ \myPhi(x_{1},x_{1}+x_{2},\emptyset,y_{1})} 
  \myShifted[1]{y_{1}}{ \myPhi(x_{1},\myset{X}+x_{1},\emptyset,\myset{Y}+y_{2})} 
  = 
  \myPhi(x_{1},\myset{X}+x_{1}+x_{2},\emptyset,\myset{Y}+y_{1}+y_{2}) 
\end{equation}
leads to the recurrence 
\begin{equation}
\label{rec:zax}
\begin{array}{ll}
  \mathfrak{o}_{\myset{X}+x_{1}+x_{2},\myset{Y}+y_{1}+y_{2}}
  & = 
  \myShifted[1]{\myset{X}}{ \myPhi(x_{1},x_{1}+x_{2},\emptyset,y_{1})} 
  \myShifted[1]{y_{1}}{\mathfrak{o}_{\myset{X}+x_{1},\myset{Y}+y_{2}} } 
\\&
  + 
  \myShifted[1]{\myset{X}}{ \myPhi(x_{2},x_{1}+x_{2},\emptyset,y_{1})} 
  \myShifted[1]{{y_{1}}}{ \mathfrak{o}_{\myset{X}+x_{2},\myset{Y}+y_{2}} }
\end{array}
\end{equation}
where $\myShifted{\myset{X}}{}$ denotes the shift 
$\vec{z} \to \vec{z} + \abel{X}$.

In the limiting case of $\myset{X} = \{ x_{1},x_{2},x_{3} \}$, 
$\myset{Y} = \{ y \}$ identity \eref{id:zax31} yields
\begin{equation}
  \mathfrak{o}_{\{x_{1},x_{2},x_{3}\},y} = 0.
\end{equation}
Thus, equation \eref{rec:zax} implies 
\begin{equation}
  \mathfrak{o}_{\myset{X},\myset{Y}} = 0, \qquad
  \mysetsize{X} = |\myset{Y}|+2 
\end{equation}
which completes the proof of proposition \ref{prop:zax}.

\section{Proof of proposition \ref{prop:zbx}. \label{proof:zbx}} 

Replacing in \eref{id:zax} $\myset{X}$ with $\myset{X} + a$, 
\begin{equation}
  \myPhi(a,\myset{X} + a,\emptyset,\myset{Y}) 
  + 
  \sum_{x \in \myset{X}} \myPhi(x,\myset{X} + a,\emptyset,\myset{Y}) = 0,
\end{equation}
and noting that 
\begin{equation}
  \myPhi(x,\myset{X} + a,\emptyset,\myset{Y}) 
  = 
  - 
  \myPhi(x,\myset{X},a,\myset{Y}+a) 
  \myPhi(a,\myset{X} + a,\emptyset,\myset{Y}) 
  \qquad
  (x \in \myset{X})
\end{equation}
one can obtain 
\begin{equation}
    1 
    -
    \sum_{x \in \myset{X}} \myPhi(x,\myset{X},a,\myset{Y} + a) 
  = 0
\end{equation}
which, after replacing $\myset{Y} + a \to \myset{Y}$, leads
to \eref{id:zbx}.
Equation \eref{id:zby} follows from \eref{id:zbx} and the symmetry 
\begin{equation}
  \myPhi(\myset{A},\myset{X},\myset{B},\myset{Y}) 
  = 
  \myPhi(\myset{B},\myset{Y},\myset{A},\myset{X}). 
\end{equation}

\section{Useful lemma. \label{proof:zcx}} 

Here we prove a useful statement that is used in what follows.

{\def\theproposition{C.1}
 
\begin{proposition} \label{prop:zcx}

For arbitrary sets $\myset{X}$, $\myset{Y}$ and their subsets 
$\myset{A} \subset \myset{X}$, $\myset{B} \subset \myset{Y}$
related by 
\begin{equation}
  \mysetsize{X} - 2\mysetsize{A} + 2
  =
  \mysetsize{Y} - 2 \mysetsize{B} 
\end{equation}
the function $\Phi$ satisfies 
\begin{equation}
\label{id:zcx}
  \sum_{x \in \myset{A}} 
  \myPhi(\myset{A}\backslash x,\myset{X},\myset{B},\myset{Y})  
  = 
  \sum_{y \in \myset{Y}\backslash\myset{B}} 
  \myPhi(\myset{A},\myset{X},\myset{B}{+}y,\myset{Y}), 
\end{equation}
\begin{equation}
\label{id:zcy}
  \sum_{x \in \myset{X}\backslash\myset{A}} 
  \myPhi(\myset{A}{+}x,\myset{X},\myset{B},\myset{Y})  
  = 
  \sum_{y \in \myset{B}} 
  \myPhi(\myset{A},\myset{X},\myset{B}\backslash y,\myset{Y}). 
\end{equation}

\end{proposition} 
}

To obtain this result, we start with \eref{id:zax}, 
replace  
$\myset{X}$ with $\myset{X}_{1} + \myset{Y}_{2}$ and
$\myset{Y}$ with $\myset{X}_{2} + \myset{Y}_{1}$, 
split the sum
\begin{equation}
  \sum_{x \in \myset{X}_{1}} 
    \myPhi( x,\myset{X}_{1} + \myset{Y}_{2},\emptyset, \myset{X}_{2} + \myset{Y}_{1} )
  + 
  \sum_{y \in \myset{Y}_{2}} 
    \myPhi( y,\myset{X}_{1} + \myset{Y}_{2},\emptyset, \myset{X}_{2} + \myset{Y}_{1} ) 
  = 0 
\end{equation}
and use the identities 
\begin{equation}
\fl\qquad
  \myPhi( x,\myset{X}_{1} + \myset{Y}_{2},\emptyset,\myset{X}_{2} + \myset{Y}_{1} ) 
  \myPhi(\myset{X}_{1},\myset{X}_{1} + \myset{X}_{2},\myset{Y}_{1},
         \myset{Y}_{1} +\myset{Y}_{2} ) 
  =
  - 
  \epsilon_{\myset{X}_{1} \myset{Y}_{1}} 
  \myPhi(\myset{X}_{1}\backslash x,\myset{X}_{1} + \myset{X}_{2}, 
         \myset{Y}_{1},\myset{Y}_{1} + \myset{Y}_{2} ) 
\end{equation}
\begin{equation}
\fl\qquad
  \myPhi( y,\myset{X}_{1} + \myset{Y}_{2},\emptyset,\myset{X}_{2} + \myset{Y}_{1} ) 
  \myPhi( \myset{X}_{1},\myset{X}_{1} + \myset{X}_{2},\myset{Y}_{1}, 
          \myset{Y}_{1} + \myset{Y}_{2} ) 
  =
  \epsilon_{\myset{X}_{1} \myset{Y}_{1}} 
  \myPhi( \myset{X}_{1},\myset{X}_{1} + \myset{X}_{2},\myset{Y}_{1} + y,
          \myset{Y}_{1} + \myset{Y}_{2} ) 
\end{equation}
where 
$ \epsilon_{\myset{X} \myset{Y}} = (-)^{ \mysetsize{X} \mysetsize{Y} } $. 
Thus, we arrive, after the substitution 
$ \myset{X}_1 \to \myset{A} $,
$ \myset{X}_2 \to \myset{X} \backslash \myset{A} $,
$ \myset{Y}_1 \to \myset{B} $ and
$ \myset{Y}_2 \to \myset{Y} \backslash \myset{B} $,
at \eref{id:zcx}.

Identity \eref{id:zcy} can be proved in a similar way or obtained from 
\eref{id:zcx} using the symmetry of the function 
$\myPhi(\myset{A},\myset{X},\myset{B},\myset{Y})$.

\section{Proof of proposition \ref{prop:ZAX}. \label{proof:ZAX}} 

For two fixed sets, $\myset{X}$ and $\myset{Y}$ related by 
$ \mysetsize{X} = \mysetsize{Y} + 2 $, consider the left-hand side of 
\eref{id:ZAX} as a function of $\myset{B}_{m}$, 
\begin{equation}
  \mathfrak{f}(\myset{B}_{m}) 
  = 
  \sum_{ \myset{A}_{m+1} \subset \myset{X} } 
  \myPhi(\myset{A}_{m+1},\myset{X},\myset{B}_{m},\myset{Y})  
\end{equation}
where, recall, subscripts $m$ and $m+1$ indicate the size of the corresponding 
sets.
 
Using a simple combinatorial identity
\begin{equation}
\label{id:comb}
  \sum_{\myset{A}_{m+1} \subset \myset{X}} 
  f(\myset{A}_{m+1})
  = 
  \frac{1}{m+1}
  \sum_{\myset{A}_{m} \subset \myset{X}} 
  \sum_{x \in \myset{X} \backslash \myset{A}_{m}} 
  f(\myset{A}_{m}{+}x)
\end{equation}
one can present $\mathfrak{f}(\myset{B}_{m})$ as 
\begin{equation}
  \mathfrak{f}(\myset{B}_{m}) 
  = 
  \frac{1}{m+1}
  \sum_{\myset{A}_{m} \subset \myset{X}} 
  \sum_{x \in \myset{X} \backslash \myset{A}_{m}} 
  \myPhi(\myset{A}_{m}{+}x,\myset{X},\myset{B}_{m},\myset{Y}) 
\end{equation}
which, by virtue of \eref{id:zcy}, yields
\begin{eqnarray}
  \mathfrak{f}(\myset{B}_{m}) 
  & = & 
  \frac{1}{m+1} 
  \sum_{\myset{A}_{m} \subset \myset{X}} 
  \sum_{y \in \myset{B}_{m}} 
  \myPhi(\myset{A}_{m},\myset{X},\myset{B}_{m}\backslash y,\myset{Y}) 
\\
  & = & 
  \frac{1}{m+1}
  \sum_{y \in \myset{B}_{m}} 
  \mathfrak{f}(\myset{B}_{m}\backslash y) 
\\
  & = & 
  \frac{1}{m+1}
  \sum_{\myset{B}_{m-1} \subset \myset{B}_{m}} 
  \mathfrak{f}(\myset{B}_{m-1}) 
\end{eqnarray}
Thus, we have expressed $\mathfrak{f}$ on a $m$-set as a combination of 
$\mathfrak{f}$ on $(m-1)$-sets. 
In the limiting case, $\myset{B}_{0} = \emptyset$, identity \eref{id:zax} 
implies 
\begin{equation}
  \mathfrak{f}(\emptyset) 
  = 
  \sum_{x \in \myset{X}} 
  \myPhi(x,\myset{X},\emptyset,\myset{Y}) 
  = 0 
\end{equation}
which, by induction, leads
\begin{equation}
  \mathfrak{f}(\myset{B}_{m}) = 0,
  \qquad 
  m=1,2,...
\end{equation}
Adding the constraint $m \le \mysetsize{Y}$, one arrives at the statement of 
proposition \ref{prop:ZAX}.

\section{Proof of proposition \ref{prop:ZBX}. \label{proof:ZBX}} 

As in the proof of proposition \ref{prop:ZAX}, 
for two fixed sets, $\myset{X}$ and $\myset{Y}$ related this time by 
$\mysetsize{X} = \mysetsize{Y}$, consider the left-hand side of 
\eref{id:ZBX} as a function of $\myset{B}_{m}$, 
\begin{equation}
  F(\myset{B}_{m}) 
  = 
  \sum_{ \myset{A}_{m} \subset \myset{X} } 
  \myPhi(\myset{A}_{m},\myset{X},\myset{B}_{m},\myset{Y})  
\end{equation}
Using, again, identity \eref{id:comb} one can present $F(\myset{B}_{m+1})$ as 
\begin{eqnarray}
  F(\myset{B}_{m+1}) 
  & = & 
  \sum_{\myset{A}_{m+1} \subset \myset{X}} 
  \myPhi(\myset{A}_{m+1},\myset{X},\myset{B}_{m+1},\myset{Y}) 
\\
  & = & 
  \frac{1}{m+1} 
  \sum_{\myset{A}_{m} \subset \myset{X}} 
  \sum_{x \in \myset{X} \backslash \myset{A}_{m}} 
  \myPhi(\myset{A}_{m}{+}x,\myset{X},\myset{B}_{m+1},\myset{Y}) 
\end{eqnarray}
which, by virtue of \eref{id:zcy}, yields
\begin{eqnarray}
  F(\myset{B}_{m+1}) 
  & = & 
  \frac{1}{m+1} 
  \sum_{\myset{A}_{m} \subset \myset{X}} 
  \sum_{y \in \myset{B}_{m+1}} 
  \myPhi(\myset{A}_{m},\myset{X},\myset{B}_{m+1}\backslash y,\myset{Y}) 
\\
  & = & 
  \frac{1}{m+1}
  \sum_{y \in \myset{B}_{m+1}} 
  F(\myset{B}_{m+1}\backslash y) 
\\
  & = & 
  \frac{1}{m+1}
  \sum_{\myset{B}_{m} \subset \myset{B}_{m+1}} 
  F(\myset{B}_{m}). 
\end{eqnarray}
Thus, we have obtained the recurrence from $m$-sets to $(m+1)$-sets. 
In the limiting case, $m=1$ or $\myset{B}_{1} = \{ y \}$, 
identity \eref{id:zbx} implies 
\begin{equation}
  F(y) 
  = 
  \sum_{x \in \myset{X}} 
  \myPhi(x,\myset{X},y,\myset{Y}) 
  = 1 
\end{equation}
which, by induction, leads to 
\begin{equation}
  F(\myset{B}_{m}) = 1,
  \qquad 
  m=2,...
\end{equation}
and hence, after adding the constraint $m \le \mysetsize{Y}$, 
to the statement of proposition \ref{prop:ZBX}.

%


\end{document}